\documentstyle[prl,aps,multicol,epsf,preprint]{revtex}
\tighten
\input epsf

\newcommand{\D}{{\rm d}}
\newcommand{\I}{{\rm i}}

\begin{document}
\draft
\title{On localization effects in underdoped cuprates}
\author{C. Castellani, P. Schwab, and M. Grilli}
\address{ Istituto di Fisica della Materia e Dipartimento di Fisica,
 Universit\`a ``La Sapienza'', 
 piazzale A. Moro 2, 00185  Roma - Italy}
\date{\today}
\maketitle
\begin{abstract}
We comment on transport experiments in underdoped LaSrCuO in the non-superconducting
phase. The temperature dependence of the resistance strongly resembles
what is expected from standard localization theory.
However this theory fails, when comparing with experiments in more detail. 
%We argue that anomalous localization effects are to be
%expected in a disordered stripe phase.
\end{abstract} 
% \pacs{PACS numbers: }

%\begin{multicols}{2}
\section{introduction}
It is generally believed that the understanding of the anomalous normal state 
properties would shed light on the understanding of the pairing mechanism of 
high-$T_c$ superconductors. In particular, low temperature transport 
experiments in the normal state should provide valuable informations on the 
physical mechanisms acting in high-$T_c$ materials.
The normal state is usually inaccessible at low temperature, due to the onset of
superconductivity. However, superconductivity  can be suppressed down to 
lowest temperatures  by applying a sufficiently high magnetic field.

In this paper we shall discuss the low temperature resistance in underdoped 
$\rm La_{2-x} Sr_x Cu O_4$ (LSCO) at high magnetic fields, where insulating 
behavior with a typical $\log T$ dependence of the resistance has been 
observed\cite{Ando95}. Various proposals have been made for the origin of 
this behavior. Anderson et al.\cite{Anderson96} 
interpreted the data in the framework of Luttinger liquid transport theory,
which predicts a power law in the c-axis and in the in-plane conductivity 
$\sigma_{\rm c} \propto \sigma_{\rm ab} \propto T^{2\alpha} $. 
Although it is not unreasonable to fit
the experiments by a power law, much better fits are achieved assuming the
logarithmic behavior.
Alexandrov\cite{Alexandrov96} reported a logarithmically divergent resistivity
in a bipolaron model in presence of disorder, however
this model has been repeatedly questioned\cite{Chakraverty98}.
Varma\cite{Varma97} argued that in a non Fermi-liquid even small disorder drives
the density of states to zero and thus drives the system to an insulator.

Standard localization effects have been discussed as source of the increasing 
resistance since the first experiments\cite{Preyer91,Ando95}. 
Evidence against this interpretation has arised from measurements of the Hall 
resistance\cite{Ando97}, which is nearly temperature independent.
However, since the mechanism which dominates the Hall effect in the cuprates 
is not clear, it is hard to make conclusions. So far a detailed analysis of 
the temperature dependence of the resistivity versus the predictions of 
localization theory has not been given. 
This will be subject of the present paper.

In the next section we shall briefly recall standard localization theory in 
two dimensional systems, discussing
both coherent backscattering and interaction effects.
Then we shall apply localization theory to LSCO. We will demonstrate that 
the cross-over from 3d to 2d localization in LSCO is expected 
near optimal doping. However the amplitude of the increase of resistance in 
that material in the underdoped region does not fully agree with standard 
2d localization theory. We will argue that anomalous localization effects
are indeed to be expected in that region in the presence of a disordered 
stripe phase. 
 
\section{Localization in 2d systems}
In two dimensions arbitrarily weak disorder can
localize all electronic states. This famous result follows from the single 
parameter scaling theory of localization\cite{scaling79}. 
This theory is justified, when interactions are negligible.
In the weak disorder limit ($k_F l \gg 1$) the conductivity at low temperature is given by 
\begin{equation}   
\sigma = \sigma_0 -{1\over \pi} {e^2 \over h} \ln(\tau_\phi/\tau)
,\end{equation}
where
$\sigma_0$ is the `classical' Drude conductivity, and the logarithmic term
is due to quantum corrections, with 
$\tau$ the (elastic) scattering time, and $\tau_\phi$ the dephasing time. 
These ``weak localization'' corrections are due to
quantum interferences for electrons which are diffusing along paths 
containing closed loops  
(coherent backscattering). 
Dephasing is due to inelastic processes, 
with $1/\tau_\phi \propto T^p$, leading
to a correction to conductivity, which is logarithmic in temperature,
$\delta \sigma = -(e^2/h)(p/\pi) |\ln( T\tau )|$.

However, single parameter scaling fails in 
presence of interactions, where a scaling 
theory including electron interactions is needed.
Such a theory has been put forward by Finkel'stein\cite{oldtheory}.
In perturbation theory, 
new singular contributions to the conductivity are found, 
which are -- in 2d -- proportional to $\log T$.
These singular corrections to the conductivity are due to the 
interplay between disorder and interaction and arise because
on distances that are larger than the
mean free path electrons move slowly and have more time to interact
with each other.

The correction to the resistivity due to this mechanism is
\begin{equation}
{\delta \rho }
= {1\over \pi} \rho^2 \left\{ 
1 + 3 \left[ 1- {1+\gamma_2 \over\gamma_2}\ln(1+\gamma_2)\right]
\right\}  |\ln (T\tau ) |
\end{equation}
where
``$1$'' is due to interactions
in the singlet channel, and
``$3[\cdots]$'' are due to the triplet channels.
The universal value of the singlet amplitude is due to the long range
nature of the Coulomb interaction, since after screening, the dimensionless interaction
equals in the long wavelength limit always one, 
$(\D n / \D \mu) V(q)/[1+(\D n / \D \mu )V(q)]=1$. 
$\gamma_2$ is an interaction parameter which is related to the Landau 
parameter $-A^0_a$.

Analyzing the renormalization group flow, a metal insulator 
transition has been found, and a phase with finite resistance at
zero temperature exists\cite{oldtheory,Castellani98}.
The interacting system avoids localization, since the triplet amplitude 
becomes relevant under scaling ($\gamma_2 \to \infty$). 
The disordered Fermi liquid tends to form ferromagnetic polarons 
(local moments). Experimental evidence for enhanced spin fluctuations has 
been found near the metal-insulator transition in 3d Si:P.   
Experimental prove for large spin fluctuations near the recently discovered 
metal-insulator transition in 2d MOSFETs is still lacking, 
but the effects are currently discussed. 

In the context of underdoped LSCO it is important that a magnetic field 
drastically modifies the above scenario. 
A magnetic field, besides suppressing coherent backscattering due to orbital 
effects, also suppresses the triplet $M=\pm 1$ due to Zeeman splitting. As a 
consequence, spin fluctuations remain small and $\gamma_2 \to 1$ under 
renormalization. Thus the singlet
term dominates leading to an insulating behavior as indeed seen in experiments
in the low temperature regime \cite{Ando95}. We shall argue below that an 
analogous result can be envisaged in the extreme underdoped regime even 
in the absence of magnetic field.

\section{Application to cuprates}
In standard studies of localization the magnetoresistance is a main 
probe for extracting important informations on both backscattering and 
interaction effects. Unfortunately, in cuprates many different 
contributions to the magnetoresistance have been observed related to 
the superconducting fluctuations \cite{Varlamov} and to the peculiar 
behavior of the Hall conductance \cite{Tyler}. These effects may mask the 
localization contributions and thus make an interpretation difficult. 
Therefore we concentrate in the following on the temperature dependence of 
the resistance under conditions, where possibly the above complications are 
not present. 

We refer to two types of experiments.
Extremely underdoped LSCO ($x <0.04$), which is non-superconducting even 
in absence of magnetic fields, and
the system near optimum doping, but large magnetic field ($0.04 <x \leq 0.16$).
In both cases
$\log T$ behavior in the resistivity or conductivity has been observed, 
suggesting that the physics of disorder and interaction 
in two dimensions is relevant.
However, despite a substantial anisotropy, the LSCO materials
are bulk systems of weakly coupled layers. This raises the
relevant issue of the effective dimensionality  
of LSCO with respect to localization.
Weak localization in a nearly two dimensional metal has been considered by
Abrikosov\cite{Abrikosov}.
We performed a similar calculation for the interaction contribution. We generalized the
model of c-axis transport of Ref.\cite{Abrikosov} incorporating interplanar 
disorder as discussed in Ref.\cite{Rojo}.
We found that the system behaves two-dimensional, when
the tunneling time $\tau_{\rm tun} $ between layers is larger 
than the time of the slowest processes which are contributing to localization. 
For weak localization, the relevant time scale is the phase coherence time 
$\tau_\phi$, whereas the time scale for the interaction contribution, which 
is the relevant one in high magnetic field, is $\hbar/T$. Therefore the cross 
over from 3d to 2d is defined by 
$1/\tau_{\rm tun} \sim {\rm max} (\, T/\hbar, 1/\tau_{\phi})$.    
The tunneling rate is hard to estimate directly, since it is not 
clear if processes which conserve in-plane momentum dominate,  
$\hbar/\tau_{\rm tun} = 2|t_\perp|^2 \tau/\hbar $,
or momentum non-conserving processes dominate, for which
$\hbar/\tau_{\rm tun} = 2\pi |V|^2 N(0)$.
Here $t_\perp$ and $V$ are tunneling amplitudes, $\tau$ the quasiparticle
lifetime and $N(0)$ is the density of states. More conveniently,
the tunneling rate is determined from the c-conductivity,
since $\sigma_c = 2e^2 (1/\tau_{\rm tun})N(0) c$.
Inserting the free electron value for the 2d density of states with $m^{*}/m 
\simeq 3$  and 
$c=6.5${\AA} in LSCO we
find for $\rho_c = 3\cdot 10^{-2} \Omega$cm 
a tunneling rate of $\hbar/\tau_{\rm tun} \simeq 20$K.
In Ref.\cite{Ando95} the c-resistivity at 20K was between 
$\rho_c \simeq 2\cdot 10^{-2} \Omega$cm ($x=0.22$) and
$\rho_c \simeq 3\Omega$cm ($x=0.08$).
By comparing $\hbar/\tau_{\rm tun}$ to the temperature at $20$K, we conclude
that the samples near optimal doping  at $ x \simeq .17$ 
are near the dimensional  
cross-over from 3d to 2d. 
The underdoped samples are presumably still two-dimensional at $20$K, whereas
the overdoped samples may be three-dimensional. 
In this case a $\sqrt{ T}$ correction to the conductivity at low temperature
instead of $\log T$ is expected.  

Preyer et al. \cite{Preyer91} reported the conductivity in highly underdoped
$\rm La_{2-x}Sr_xCuO_4$.
For a sample with $x=0.04$ they found a logarithmic correction 
to the conductivity
\begin{equation}
\sigma = \sigma_0 - {\lambda \over \pi} {e^2\over h} |\ln(T/T_0) |
\end{equation} 
with $\lambda \approx 0.7$ over more than a decade of temperature, 
from $\sigma \approx 0.5 e^2/h$ down to 
$\sigma \approx 0.1 e^2/h$, between $\sim 100$K and $\sim 10$K,
where a crossover to variable range hopping was observed.
This seems to be in good agreement with standard localization theory. 
The magnetoresistance is negative and isotropic. Assuming that weak 
localization is relevant, the isotropic magnetoresistance is 
not consistent with conventional theory.
On the other hand, conventional theory is build for a non-magnetic 
Fermi liquid, whereas here a theory in a doped anti-ferromagnet is needed, 
with peculiar quasi-particles (hole pockets). 
Also in absence of long range order conventional theory has to be modified.
A short magnetic correlation length makes the system similar to a spin-glass.
The random magnetic field in a spin-glass is assumed to suppress quantum 
corrections to the conductivity from coherent  backscattering and from 
interactions in the triplet channels. Quantum corrections in the remaining 
interaction singlet channel are
field independent, which suggests that the experimentally seen 
magnetoresistance is due to a different mechanism.
Although a full theory of localization in a doped anti-ferromagnet is lacking,
the presence of a $\log T$ of the correct magnitude, and the correct scale of
the cross-over to strong localization strongly suggests that the physics of
disorder and interactions is here relevant.
 
Ando et al.\cite{Ando95} studied LSCO for higher doping, 
suppressing superconductivity by strong magnetic
fields with pulses of up to $60$T.  
There is practically no magnetoresistance in the normal 
state, i.e. once superconductivity
is destroyed the resistance saturates.
Below $x\approx 0.16$ they found an insulating behavior
at low temperature, i.e.\ $\delta\rho/\delta T<0$. Both in ab and c direction a 
$\log T$ in the
resistance was found,
\begin{equation}
\rho_{\rm ab}= \rho^0_{\rm ab} + \alpha | \ln T/T_0 |
\,\, {\rm and} \,\,\,\,
\rho_{\rm c} = \rho^0_{\rm c}  + \beta  | \ln T/T_0 | 
,\end{equation}
where $\alpha $ and $\beta$ are sample dependent. Analyzing the 
amplitude of the  
$\log T$ near the onset ($\approx 20$K for $\rho_{\rm ab}$), i.e. 
calculating the ``interaction
constant'' $\lambda_{\rm exp}$ according to
\begin{equation} \label{eq6}
{1\over \rho_{\rm ab}} {\delta\rho_{\rm ab} \over \delta| \ln T |} = 
{\lambda_{\rm exp} \over \pi}
{\rho_{\rm ab} \over h/e^2 } 
,\end{equation}
we found from experimental data of Ref.\cite{Ando95} $\lambda_{\rm  exp} 
\approx 2$--5.
Standard localization theory predicts in high magnetic field
\begin{equation}
\lambda_{\rm theo} = 1+ \left[ 1- {1+\gamma_2\over \gamma_2} \ln(1+\gamma_2) 
\right]
\end{equation}
which is of order one, but never larger than two,
since stability of the Fermi liquid requires $\gamma_2 > -1$.
Apparently $\lambda_{\rm exp}$ and
$\lambda_{\rm theo}$ are of the same order of magnitude.

To the first view the experiments seem to be in reasonable 
agreement with theory.
There are, however, a number of problems:
(a) While theory predicts $\log T$ in the conductivity, it is 
experimentally seen in a large range
of resistivity.
(b) The ratio $\rho_{\rm ab}/\rho_{\rm c}$ does not depend on temperature, 
i.e. the $\beta$-function in ab- and c-direction is the same,
$\delta \ln \rho_{\rm c} / \delta \ln T = \delta \ln \rho_{\rm ab}/ \delta \ln T$.
This is predicted for anisotropic, but three dimensional localization\cite{Wolfle84}.
In the temperature region of two dimensional localization a logarithmic correction
to the c-conductivity is expected due to the corrections to the tunneling density
of states, $N$. Explicitly working out the theory, we found
$\delta \ln \rho_c / \delta\ln T = 2 \delta \ln N / \delta \ln T$, which in general
differs from the correction to the resistivity in ab-direction.
(c) A third problem arises from a quantitative analysis of the amplitude
of the $\log T$, which is of the right order of magnitude, but nevertheless is
too large.

Further investigating problem (c), we found an intriguing relation between 
the experimentally measured amplitude of the log-corrections 
$\lambda_{\rm exp}$ and the amount of
disorder as obtained from the absolute value of resistivity at 
some fixed temperature $\rho(T=T_0)\equiv \rho_0$.
In Tab.I
%\ref{tab1} 
we report $\lambda_{\rm exp}$ for a number of 
samples, comparing ``clean'' (low $\rho_0$) and ``dirty'' samples of the 
same material and dopant concentration\cite{Ando95,Ando97,Jing91}.
Whereas $\lambda_{\rm theo}$ is independent of disorder, 
the experimentally determined
value decreases with increasing
dirtiness: As shown in the table the product 
$\rho_0 \lambda_{\exp}$ is nearly
disorder independent. Moreover, $\rho_0 \lambda_{\rm exp}$ seems
to decrease with increasing doping. For the four LSCO samples we 
report in Tab.I
%\ref{tab1}
the product $\rho_0 \lambda_{\rm exp} x^2$ is roughly 
independent from disorder and doping.

A second observation is, that various features of the
anomalous localization can be described phenomenologically
by using the Drude formula for the conductivity and taking 
the scattering rate from the ansatz
\begin{equation}
{1\over \tau} = {1\over \tau_0 } \left( 1+ a|\ln T/T_0| \right).
\label{obs2}
\end{equation}
This logarithmically enhanced scattering rate appears 
directly in the resistivity,
not in the conductivity, and is therefore consistent
with the property outlined in point (a) above. 
Moreover the constant ratio of $\rho_{\rm ab}/\rho_{\rm c}$ 
can be reconciled with a $\log T$ correction which is typical of two dimensional
systems (problem (b) outlined above)
by the assumption (\ref{obs2}) if 
tunneling between planes is dominated by  
momentum conserving processes.
Finally, if dirtiness only affects $1/\tau_0$, but not the logarithmic term, 
the amplitude of the $\log  T$ as a function
of disorder behaves according to the experimental observation
discussed above, $\rho_0 \lambda_{\rm exp} = \rm const$.
 
\section{Discussion}
We discussed some features of transport experiments in LSCO compounds
in the normal state. On the one hand the $\log T$ correction in strongly 
underdoped LSCO ($x\approx 0.04$) appears to be consistent with 
standard localization theory, although 
an explanation for the magnetoresistance is still lacking. On the other hand, 
for higher doping, $0.04 \lesssim x \lesssim 0.16$, 
standard theory is not able to explain the experiments.

There are several reasons why the conventional
``old fashioned'' localization theory is not expected to work well in 
strongly correlated anisotropic systems like the cuprates.
One first possibility is that the cuprates can not be described by
the Fermi liquid theory as already pointed out in the introduction 
\cite{Anderson87}.
If this is the case, a new localization theory starting from
a clean non-Fermi liquid system should be devised 
\cite{Anderson96,Varma97}. Alternatively, 
a singular interaction could be responsible for both the disruption of
the Fermi liquid and of the anomalous localization effects.
Mirlin and W\"olfle\cite{Mirlin97} reported anomalous localization effects 
within a gauge field theory\cite{gaugefield}, where particles interact via a singular 
transverse gauge field propagator $\propto 1/(-\I \omega \sigma + \chi q^2)$. 
At low temperature a $\log T$ correction has been found, with an
amplitude that depends on 
resistance itself, $\lambda \propto \ln(1/\rho)$.

In the quantum critical point (QCP) scenario of high $T_c$ superconductivity, 
a QCP exists near optimum doping, with an ``ordered'' stripe phase 
in the underdoped regime \cite{QCP,stripes,Varma96}. In this context,
possible sources of singular interactions are soft modes from 
dynamical stripes, or critical fluctuations near the  QCP.
Specifically, the interaction near the stripe critical wave-vector ${\bf q}_c$ 
may be parameterized as\cite{QCP}
\begin{equation}
\Gamma( {\bf q }, \omega ) \approx - {A \over \omega_{\bf q} -\I \omega }, 
\end{equation} 
where
$\omega_{\bf q} \approx B(|{\bf q - q}_c|^2 + \kappa^2 )$.
Quantum corrections to the conductivity due to exchange of 
these fluctuations are,
to first order in $\Gamma({\bf q},\omega)$, proportional 
to the Fermi-surface average of
the static interaction,
$\Gamma_{\rm QCP} = \langle \Gamma({\bf q} )\rangle_{FS} \propto - A 
\log(k_F/\kappa )$.
The interaction is attractive, leading to an increase of localization.
For large $\Gamma_{\rm QCP}$ quantum corrections have to be
calculated beyond first order. 
However standard theory \cite{oldtheory} does not apply here, due to the strong
frequency dependence of the interaction.
We have preliminary results indicating, that quantum 
($\omega_{\bf q} > T$) and classical
($\omega_{\bf q} < T$) fluctuations contribute differently to localization.

Finally one should also think at the possibility of a non-conventional
source of disorder. In particular, if a disordered nearly static stripe phase
is realized in these systems one should, besides the conventional 
impurity disorder, also consider the disorder coming from 
domain boundaries and other topological defects of the striped textures. 

An appealing possibility is that a disordered stripe phase could be 
responsible for the anomalous localization behavior both by introducing a
singular scattering between the electrons and by
providing topological disorder via domain 
boundaries. As a consequence, any mechanism like impurity or
lattice pinning of (static or slow) stripe fluctuations
is expected to reduce the amplitude of the effective interaction and 
the $\log T$ corrections, in agreement with the central observation of the 
present work that $\rho_0\lambda_{\rm exp}$ is nearly constant.
This expectation is supported by the observation, that at 
$x=1/8$, where stripes are more
ordered, the $\log T$ is less strong and the resistance saturates at low 
temperature\cite{Ando95,Castellani97}, as in 
$\rm La_{1.48}Nd_{0.4}Sr_{0.12}CuO_4$\cite{Tranquada96}.

The very speculative character of these considerations calls
for a detailed theory, which is not available at the moment.
%\acknowledgments

%\end{multicols}

\newpage
\begin{table}
\begin{tabular}{|c|c|c|c|}
$\rho_0=.26$ & $\lambda_{\rm exp} =5.2$  & LSCO, $x=0.08$ & $\rho_0\lambda_{\rm exp} = 1.3$ \\  
$\rho_0=.83$ & $\lambda_{\rm exp} =1.87$ & LSCO, $x=0.08$ & $\rho_0\lambda_{\rm exp} = 1.5$ \\ 
\hline
$\rho_0=.15$ & $\lambda_{\rm exp} =4.2$ & LSCO, $x=0.13$ & $\rho_0\lambda_{\rm exp} = 0.63$ \\  
$\rho_0=.20$ & $\lambda_{\rm exp} =3.1$ & LSCO, $x=0.13$ & $\rho_0\lambda_{\rm exp} = 0.60$ \\
\hline
$\rho_0=.046$ & $\lambda_{\rm exp} =2.5$ & Bi2201, $x=?$& $\rho_0\lambda_{\rm exp} = 0.115$ \\
$\rho_0=0.161$& $\lambda_{\rm exp} =0.6$ & Bi2201, $x=?$& $\rho_0\lambda_{\rm exp} = 0.10$ 
\end{tabular}
\label{tab1}
\vspace{1cm}
\caption{Interaction ``constant'' $\lambda_{\rm exp}$ as determined from
clean and dirty LSCO
\protect\cite{Ando95,Ando97} 
and Bi2201 
\protect\cite{Ando97,Jing91}, 
see Eq.\ (\ref{eq6}). The dopant concentration of the Bi compounds, 
i.e. the concentration of holes in the CuO planes, is not clear to us, 
however the samples are assumed to be overdoped. 
$\rho_0$ is the sheet resistance in 
units of $h/e^2$ at $T\approx 20$K. Note that the product
$\rho_0 \lambda_{\rm exp}$ assumes the
same value for clean and dirty samples.}
\end{table}


\begin{references}

\bibitem{Ando95}       Y. Ando, G.S. Boebinger, A. Passner, T. Kimura and K. Kishio,
                       Phys. Rev. Lett. {\bf 75}, 4662 (1995);
                       G.S. Boebinger et al., 
                       Phys. Rev. Lett. {\bf 77}, 5417 (1996).
\bibitem{Anderson96}   P.W. Anderson, T.V. Ramakrishnan, S. Strong, and D.G. Clarke,
                       Phys. Rev. Lett. {\bf 77}, 4241 (1996).
\bibitem{Alexandrov96} A.S. Alexandrov, preprint, cond-mat/9610065.
\bibitem{Chakraverty98}B.K. Chakraverty, J. Ranninger, and D. Feinberg,
                       Phys. Rev. Lett. {\bf 81}, 433 (1998);
                       A.S. Alexandrov, preprint, cond-mat/9807185 (comment).
\bibitem{Varma97}      C.M. Varma,
                       Phys. Rev. Lett. {\bf 79}, 1535 (1997).
\bibitem{Preyer91}     N.W. Preyer, M.A. Kastner, C.Y. Chen, R.J. Birgenau, 
                       and Y. Hikada
                       Phys. Rev. B {\bf 44}, 407 (1991);
                       B. Keimer et al,
                       Phys. Rev. B {\bf 46}, 14034 (1992).
\bibitem{Ando97}       Y. Ando et al.,
                       Phys. Rev. B {\bf 56}, 8530 (1997).
\bibitem{scaling79}    E. Abrahams, P.W. Anderson, D.C. Licciardello, and
                       T.V. Ramakrishnan Phys. Rev. Lett. {\bf 42}, 673 (1979).
\bibitem{oldtheory}    A.M. Finkel'stein JETP {\bf 57}, 97 (1983);
                       C. Castellani, C. Di Castro, P.A. Lee, and M. Ma,
                       Phys. Rev. B {\bf 30}, 527 (1984).
                       For a review see: P.A. Lee and T.V. Ramakrishnan,
                       Rev. Mod. Phys. {\bf 57}, 787 (1985);
                       D. Belitz and T.R. Kirkpatrick, Rev. Mod. Phys.
                       {\bf 66}, 261 (1994).
\bibitem{Castellani98} C. Castellani, C. Di Castro, and P.A. Lee, 
                       Phys. Rev. B {\bf 57}, 9381 (1998).
\bibitem{Varlamov}     A. Varlamov et al., 
                       Adv. in Phys., in press (1998)
\bibitem{Tyler}        A.W. Tyler et al., 
                       Phys. Rev. B {\bf 57}, 728 (1998). 
\bibitem{Abrikosov}    A.A. Abrikosov,
                       Phys. Rev. B {\bf 50} 1415 (1994).
\bibitem{Rojo}         A.G. Rojo and K. Levin,
                       Phys. Rev. B {\bf 48} 16861 (1993).
\bibitem{Wolfle84}     P. W\"olfle and R.N. Bhatt,
                       Phys. Rev. B {\bf 30}, 3542 (1984);
                       R.N. Bhatt, P. W\"olfle, and T.V. Ramakrishnan,
                       Phys. Rev. B {\bf 32}, 569 (1985).
\bibitem{Jing91}       T.W. Jing, N.P. Ong, T.V. Ramakrishnan,
                       J.M. Tarascon, and K. Remschnig,
                       Phys. Rev. Lett. {\bf 67}, 761 (1991).
\bibitem{Anderson87}   P.W. Anderson, Science {\bf 235}, 1196 (1987);
                       Phys. Rev. Lett. {\bf 64}, 1839 (1990); {\bf 65}, 2306 (1990).
\bibitem{Mirlin97}     A. Mirlin and P. W\"olfle,
                       Phys. Rev. B {\bf 55}, 5141 (1997).
\bibitem{gaugefield}   N. Nagaosa and P.A. Lee, 
                       Phys. Rev. Lett. {\bf 64}, 2450 (1990);
                       P. A. Lee and N. Nagaosa, 
                       Phys. Rev. B {\bf 46}, 5621 (1992);
\bibitem{QCP}          C. Castellani, C. Di Castro, and M. Grilli,
                       Phys. Rev. Lett. {\bf 75}, 4650 (1995); A. Perali, 
                       C. Castellani, C. Di Castro, and M. Grilli, Phys. Rev. 
                       B {\bf 54}, 16216 (1996).
\bibitem{stripes}For experimental evidence of a stripe phase in the cuprates
                 see the contributions to this conference by 
                 A. Bianconi and by J.M. Tranquada.
                 For theoretical aspects see instead the contributions by 
                 C. Di Castro, V.J. Emery, and J. Zaanen.
\bibitem{Varma96}      Different QCPs have been suggested by 
                       C.M. Varma, Physica C {\bf 263} 39 (1996);
                       P. Montoux and D. Pines, 
                       Phys. Rev. B {\bf 50}, 16015 (1994) and references therein.
\bibitem{Castellani97} C. Castellani, C. Di Castro and M. Grilli,
                       cond-mat/9709278.
\bibitem{Tranquada96}  J.M. Tranquada, J.D. Axe, N. Ichikawa, Y. Nakamura, 
                       S. Uchida, and B. Nachumi,
                       Phys. Rev. B {\bf 54}, 7489 (1996). 
\end{references}
\end{document}